# Spatio-Temporal Data Correlation with Adaptive Strategies in Wireless Sensor Networks

Jyotirmoy Karjee
Department of Electronic Systems Engineering
Indian Institute of Science
Bangalore, India
E-mail: kjyotirmoy@cedt.iisc.ernet.in

H.S Jamadagni
Department of Electronic Systems Engineering
Indian Institute of Science
Bangalore, India
E-mail: hsjam@cedt.iisc.ernet.in

***Abstract*—One of the major task of sensor nodes in wireless sensor networks is to transmit a subset of sensor readings to the sink node estimating a desired *data accuracy*. Therefore in this paper, we propose an accuracy model using Steepest Decent method called Adaptive Data Accuracy (ADA) model which doesn't require any a priori information of input signal statistics to select an optimal set of sensor nodes in the network. Moreover we develop another model using LMS filter called Spatio-Temporal Data Prediction (STDP) model which captures the spatial and temporal correlation of sensing data to reduce the communication overhead under *data reduction strategies* . Finally using STDP model, we illustrate a mechanism to trace the malicious nodes in the network under extreme physical environment. Computer simulations illustrate the performance of ADA and STDP models respectively.**

## I. INTRODUCTION

Recent progress in real time distributed system has made a drastic improvement for monitoring continuous data over wireless sensor networks. Such continuous monitoring of real data applications permits to observe in both time and space. In wireless sensor network, sensor nodes are deployed both in time and space to monitor the physical phenomenon of data (e.g temperature) from the physical environment [1]. For a specific time instant, sensor nodes collect the data in space domain and transmit it to the sink node.

The major task of sensor nodes is to collect the data from the physical environment. Since the data collected by the sensor nodes are generally spatially correlated [2] among them, the sensor nodes need not require transmitting all the sensor readings to the sink node. Instead a subset of sensor reading is sufficient to transmit to the sink node maintaining a desired *accuracy*. Thus exploring spatio-temporal correlated data to transmit a subset of sensor reading maintaining a desired *estimated data accuracy* at the sink node is an emerging topic in wireless sensor network and is the key interest of this paper. This procedure can reduce a significant communication overhead and energy consumption in the network. To achieve this goal, lot of data accuracy models [3],[4],[5],[6],[7],[8],[9],[10], [11] are proposed . These types of traditional accuracy model requires a priori knowledge of statistical data (e.g exact variances, covariance's) of the physical environment .Moreover these a priori knowledge of statistical data are generally not available in practice in real life scenario. Hence to the best understanding of authors, this is the first time ,we propose a model called Adaptive Data Accuracy (ADA) model which doesn't require any a priori information of statistical data of the environment. ADA model estimates a desired accuracy using adaptive Steepest -Decent method [15] at the sink node from the subset of sensor readings of the sensor nodes in the network.

Since our key interest is to transmit a subset of sensor readings from the sensor nodes, we aim to explore *data reduction strategies* [25]. In data reduction strategies, we use adaptive LMS filter to reduce the amount of data transmitted by each sensor nodes under spatially correlated data in the sensing region. Under data reduction strategies , we propose a model called Spatio-Temporal Data Prediction (STDP) model which not only reduces the communication overhead but also have the learning and tracking capability to trace the internal variations of the statistical signal in the network. STDP model uses adaptive LMS filter both at the sensor nodes and the sink node. In our model, filter at the sink node does the *joint prediction* [25] to capture the spatial and temporal data correlation among the sensor nodes in the sensing region.

Hence in STDP model, using LMS filter sink node estimates a global weighted vector which captures the spatial and temporal data correlation among all the sensor nodes in the wireless sensor region. Global weighted vector at the sink node gives the information about the statistics of data in the network. Thus global weighted vector calculated at the sink node in STDP model estimates good data at the sink node for the network. But in other prediction models [23, 24, 25], sink node doesn't capture spatial and temporal data correlation features in the sensor region since a single weighted vector is considered individually for the respective sensor node. Weighted vector has to depend on a single node data collection. If the node collects bad data then the estimated value of weighted vector degrades at the sink node. But STDP model is not restricted to predict the data from a single node like other prediction models. In STDP model, sink node has the knowledge of the statistics of whole data (spatio-temporal data correlation) of all sensor nodes using a global weighted vector. Thus STDP model does the *joint prediction* scheme at the sink node for data reduction which lags in literature [25]. In literature [25], filters at the sink node and the sensor nodes are always active. If the filter at the sensor node is always active, it consumes energy still it doesn't perform any transmission of data. But in STDP model, we have switching mode (like ON/OFF) mechanism to make the filters at sensor nodes and the sink node to be idle. Thus our model can save more energy than the existing model. In literature [26], a spatio-temporal model is illustrated where historical sensed data is taken to estimate sensor readings in current period. But in STDP model such historical data is not taken to estimate readings in current time period. In STDP model, spatio-temporal data is refreshed in a cyclic order after certain interval of time using a new global weighted vector calculated at the sink node.

Maximizing the network life time subjected to event constraint and information gathering to maximize the network life time subjected to energy constraints [18],[19] are discussed without verifying the data accuracy. Verifying data accuracy is essential before data aggregation as it degrades the accuracy level if some of the sensor nodes gets malicious [7],[12] due to extreme physical environment like heavy rain fall etc. Thus inaccurate data gets aggregated with the other correct data at the sink nodes which results incorrect data aggregation at the sink node. In literature [7], authors perform data accuracy under malicious nodes but don't incorporate to trace the number of malicious nodes in the network. Hence in this paper we propose a mechanism to find the number of malicious nodes in the network if any.

The rest of the paper is given as follows. In Section II, we explore the motivation and problem definitions of our work. In Section III, we explain briefly purpose of our system model. In Section IV, we perform the simulation as well as validation and finally conclude our work in Section V.

## II. Overview of Approach and Problem Definations

The purpose and motivation of this paper is explained in threefold which are as follows:

- In wireless sensor networks, sensor nodes senses data from the physical environment and transmit it to the sink node. In literatures, [7]-[11] we develop various data accuracy models which senses more accurate data readings from the physical environment. In Figure 1, we summarizes some of the data accuracy models for ten sensor nodes which can sense accurate data readings from the environment .These accuracy models are the traditional methods for collecting accurate data from the physical environment where variance of the sensed data is assumed to be known (e.g variances,covariances). In other words, we have to know a priori knowledge of input signal statistics of the environment. Hence in Figure 1, the signal variances and the noise variances for the input statistical signals are assumed to be known. But in this paper, we develop a model called Adaptive Data Accuracy (ADA) model which doesn't require any a priori knowledge of input signal statistics of the environment. ADA model has the capability for tracing continuous data stream for a regular time interval to estimate the required signal.

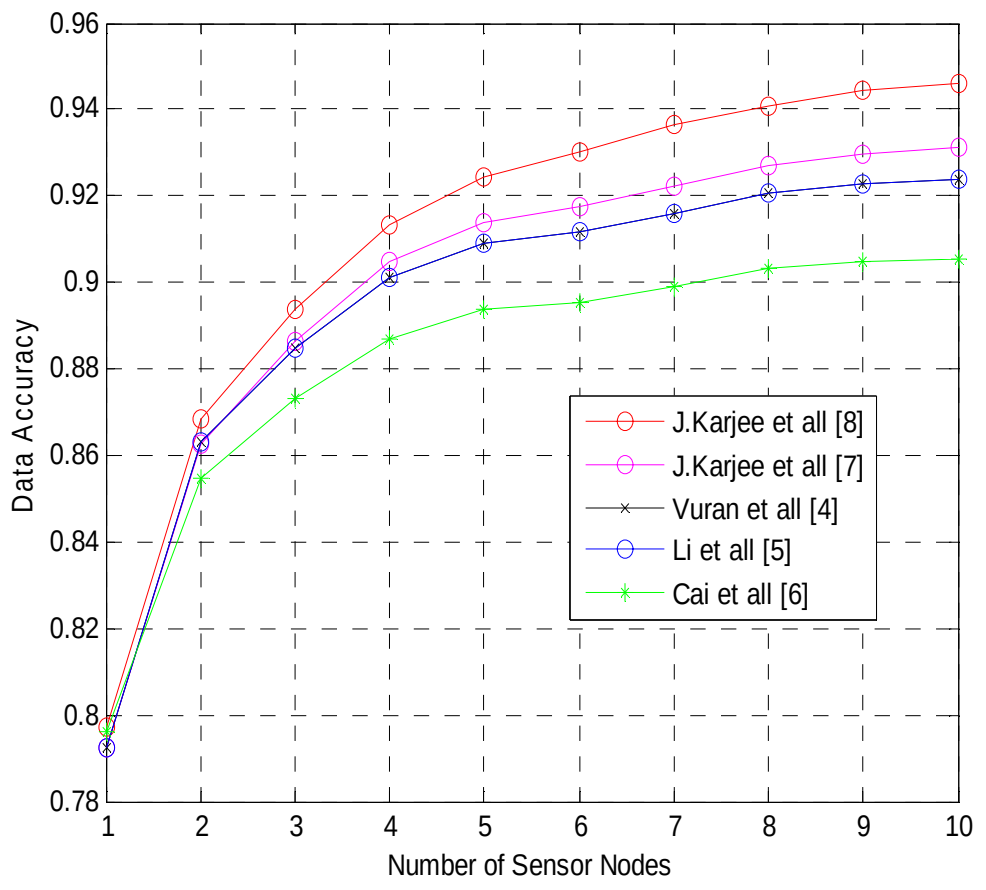

Fig.1.Data accuracy models with known input signal statistics (traditional approach)

- A Spatio-Temporal Data Prediction (STDP) model is developed to capture the signal statistics when the data are correlated among the sensor nodes in the wireless sensor network. This STDP model performs a *joint prediction* scheme which can learn and track the internal variation of the signal statistics to adopt itself with the environment. It reduces the communication overhead in the wireless sensor network based on data reduction strategies.
- In literature [7], we use traditional way of estimating data accuracy model where some of the sensor nodes get malicious due to extreme physical environment e.g heavy rainfall. In traditional model, we don't have any methodology to find the malicious nodes in the network. Hence in this paper using STDP model, we illustrate a mechanism which have not only the learning as well as tracking capability but also have the capability to trace the malicious nodes in the network.

In the next section, we discuss the system model of our propose problem definition in brief.

## III. System Model

We consider $M$ sensor nodes randomly distributed over a wireless sensor network. When a query is requested from the sink node to the sensor nodes, the sensor nodes start sensing the physical phenomenon of data e.g temperature from the environment and transmits data to the sink node. For the simplicity of our system model, we called these sensor nodes as clients and the sink node as the central server. Hence clients and server performs the following roles in wireless sensor networks for data transmission.

***Clients:*** Each sensor node $i$ can sense and observe the physical phenomenon of data in the wireless sensor network.

The observation made by each sensor node $i$ to collect the continuous block [20] of data samples up to $N$ samples over a window frame of time interval $T$ is given as

$$u_i = \{u_i^1, u_i^2 \ldots . u_i^N\} \quad (1 \times N) \quad \text{where } i \in M \tag{1}$$

The corresponding scalar measurement (desired signal) done by each sensor node $i$ is given as

$$d_i = u_i w_0 + v_i \quad \text{where } i \in M \tag{2}$$

where $w_0$ is $(N \times 1)$ an initial weighted vector with unknown matrix in the client side. $v_i$ is the temporal and spatial uncorrelated white noise. Each sensor node $i$ transmits $u_i$ observation to the sink node through additive white Gaussian noise (AWGN) channel [6],[13] in the wireless sensor network.

***Server:*** Sink node restores the observed data received from all the sensor nodes in $U$ matrix and the corresponding desired signal in $d$ matrix in the network as follows

$$U = col\{u_1, u_2 \ldots\ldots\ldots u_M\} \ (M \times N) \tag{3}$$

$$d = col\{d_1, d_2 \ldots\ldots\ldots\ldots d_M\} \ (M \times 1) \tag{4}$$

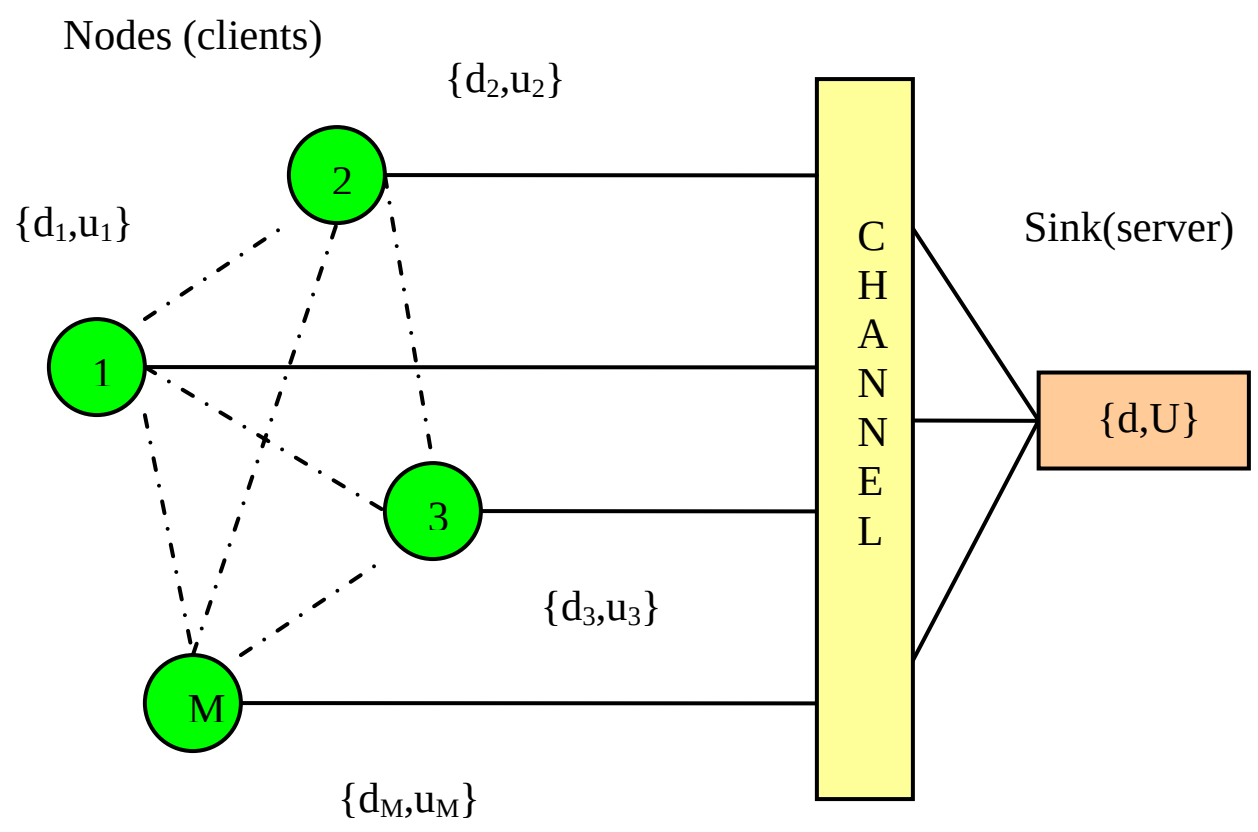

Fig 2: Architecture of System Model

In the system model, at first we construct the mathematical foundation to find the optimal sensor nodes using a model called Adaptive Data Accuracy (ADA) model. This accuracy model doesn't require a priori knowledge of input signal statistics to estimate the required signal. Secondly, we develop a Spatio-Temporal Data Prediction (STDP) model which captures the spatial and temporal correlation of data among the sensor nodes to reduce the communication overhead in the network using LMS (adaptive) filter .Finally using STDP model, we find a mechanism to trace the malicious nodes in the network .

## A. *Adaptive Data Accuracy Model*

Traditional data accuracy models require a priori and exact knowledge of input signal statistics to collect the data from the physical environment. Hence we propose a model called Adaptive Data Accuracy (ADA) model which doesn't require a priori knowledge of signal statistics and can trace the continues data stream for a regular interval of time under spatially correlated data in the sensor region .

Since $M$ sensor nodes are randomly deployed, the observed data sensed by the sensor nodes are spatially correlated among them in the sensor network. Hence we can reduce the number of sensor nodes while maintaining approximately the same data accuracy level which we achieve by $M$ sensor nodes, since the observed data are spatially correlated among them. We perform Minimum Mean Square Estimation (MMSE) [14] with adaptive [21]approach at the sink node to reduce the number of sensor nodes subjected to data accuracy for the spatially correlated observed data sensed by $M$ sensor nodes in the network.

To perform this operation, each sensor nodes $i$ in the client side transmits the spatially correlated data $u_i$ according to (1) and the corresponding measured data $d_i$ according to (2) with initial weighted vector $w_0$ to the sink node. Using adaptive Steepest Decent method [15],[22],sink node find a global statistical information $\{\Re_{uu}, \Re_{du}\}$ for the spatially correlated data in the network as follows.

$$\Re_{uu} = E[UU^T] = \begin{pmatrix} E[u_1 u_1] & \cdots & E[u_1 u_M] \\ \vdots & \ddots & \vdots \\ E[u_M u_1] & \cdots & E[u_M u_M] \end{pmatrix}$$

$$= \begin{pmatrix} \sigma_{u_1}\sigma_{u_1}\rho_{u_1u_1} & \cdots & \sigma_{u_1}\sigma_{u_M}\rho_{u_1u_M} \\ \vdots & \ddots & \vdots \\ \sigma_{u_M}\sigma_{u_1}\rho_{u_Mu_1} & \cdots & \sigma_{u_M}\sigma_{u_M}\rho_{u_Mu_M} \end{pmatrix}_{(M \times M)} \tag{5}$$

Similarly

$$\Re_{du} = E[dU] = \begin{pmatrix} E[du_1] \\ E[du_2] \\ .. \\ .. \\ E[du_M \end{pmatrix} = \sigma_d \begin{pmatrix} \sigma_{u_1}\rho_{du_1} \\ \sigma_{u_2}\rho_{du_2} \\ . \\ . \\ \sigma_{u_M}\rho_{du_M} \end{pmatrix}_{(M \times 1)} \tag{6}$$

We model spatially correlated data as a Joint Gaussian Random Variable (JGRV)'s [4],[5] as follows : $E[u_i] = 0$, $E[d] = 0$; standard deviation of $u_i = \sigma_{u_i}$ , standard derivation of $d = \sigma_d$ for $i = 1, 2, \ldots\ldots . M$ . The covariance between $u_i$ and $u_j$ is $Cov[u_i u_j] = E[u_i u_j] = \sigma_{u_i}\sigma_{u_j}\rho_{u_iu_j}$ where $\rho_{u_iu_j}$ is the correlation coefficient between $u_i$ and $u_j$ for $i \neq j$ .Again $Cov[du_i] = E[du_i] = \sigma_d \sigma_{u_i} \rho_{du_i}$ where $\rho_{du_i}$ is the correlation coefficient between $d$ and $u_i$ for $i = 1,2,\ldots\ldots.. M$ . We define correlation model [4] $K_V(.)$ as $K_V(D_{i,j}) = \rho_{u_i u_j}$ where $D_{i,j} = \| u_i - u_j \|$ is the Euclidian distance between the sensor nodes $i$ and $j$ .The covariance function is non-negative and can decrease monotonically with distance $D_{i,j} = \| u_i - u_j \|$ with limiting value of 1 at $D = 0$ and of 0 at $D = \infty$ .We adopt power exponential model [16],17] in correlation model as $K_V^{PE}(D_{i,j}) = e^{-(D_{i,j}/\theta)}$ for $\theta > 0$ where $\theta$ is called as 'Range Parameter'[4],[7].Thus we get $\rho_{u_i u_j} = e^{-(D_{i,j}/\theta)}$ and $\rho_{du_i} = e^{-(D_{d,i}/\theta)}$ . Since our aim is to minimize the cost function [22] as

$$J(w) \underset{\min w}{=} E \| d - Uw \|^2 \tag{7}$$

to get an optimal solution $w_0$ using normal equation $\Re_{du} = \Re_{uu} \; w^0$ . We start from an initial guess for $w$ and derive a procedure in recursive manner until it converges to $w_0$ . Hence the weighted vector for ADA model is given as

$$w_k = w_{k-1} + \mu [\Re_{du} - \Re_{uu} w_{k-1}] \quad (\text{Mx1}) \tag{8}$$

Where $w_k$ is the weighted vector with new guess having $k$ iterations. $w_{k-1}$ is a old guess for $(k-1)$ iterations and $\mu > 0$ is a positive step size parameter [15],[22],[27] which is calculated under spatial correlation of data among sensor nodes in the network as follows

$$\Re_{uu} = V \Lambda V^T \tag{9}$$

where $\Lambda$ is the diagonal positive entries of eigen values and $V$ is a unitary matrix satisfying $VV^T = V^T V = I$ . We pick the largest eigen value $(\lambda_{\max})$ to find $\mu$ according to [15] as given as

$$0 < \mu \leq \left(2 / \lambda_{\max}\right) \tag{10}$$

Expanding (7) using (8), we get the Minimum mean square estimation (MMSE) for the ADA model given as

$$J(w) = \sigma_d^2 - \Re_{du}{}^T w - w^T \Re_{du} + w^T \Re_{uu} w \tag{11}$$

where $\Re_{uu} = E[UU^T]; \Re_{du} = E[dU]; \sigma_d^2 = E\,|\,d\,|^2$.

The normalized data accuracy at the sink node for the network is given as

$$Accuracy(w) = 1 - \frac{J(w)}{\sigma_d^2} \tag{12}$$

This normalized $Accuracy(w)$ using ADA model calculated at the sink node is used for finding the optimal number of sensor nodes in the network subjected to data accuracy. Thus ADA model doesn't require a priori knowledge of input signal statistics to calculate the data accuracy of signal at the sink node and can trace the continuous data stream for a regular interval of time in the sensor network.

### B. *Spatio-Temporal Data Prediction Model*

We explain a Spatio-Temporal Data Prediction (STDP) model which captures the spatial and temporal correlation of data to reduce the communication overhead among the sensor nodes based on *data reduction strategies*. In data reduction strategies, sensor nodes only transmit a subset of data stream to the sink node instead of transmitting the whole data stream. This reduces communication overhead in the network. STDP model performs a *joint prediction* scheme to capture the spatial correlation among sensor nodes to reduce the communication overhead. Moreover our approach doesn't require any a priori knowledge of input signal statistics and have the learning as well as tracking capability to trace the internal variation of the signal statistics.

When a query is requested from the sink node (server) to all the sensor nodes (clients), STDP model starts transmitting the data to the sink node for *joint prediction* of data in the network as follows:

***Phase I*** *Client:* Each sensor node $i$ in the client side transmits the spatially correlated data $u_i$ according to (1) and the corresponding measured data $d_i$ according to (2) (along with initial weighted vector $w_0^{\text{int}}$ ) to the sink node. $w_0^{\text{int}}$ is an unknown vector $w_0^{\text{int}} = col\{1,1,\ldots\ldots\ldots,1\} / \sqrt{N}$ [22 ]. Initially at this moment, the filter at each sensor node $i$ and the filter to be used at the sink node are kept ideal in the network.

***Phase II*** *Server:* Sink node store the received $u_i$ observation transmitted from $i$ sensor nodes in $U$ matrix and $d_i$ in $d$ matrix according to (3) and (4) respectively. Using adaptive Steepest-decent method [15],[22] , sink node find another global statistical information $\{\mathbb{R}_{UU}, \mathbb{R}_{DU}\}$ for the spatially correlated data in the network as follows

$$\mathbb{R}_{UU} = E[U^T U] \;\; (N \times N) \text{ and } \mathbb{R}_{DU} = [U^T d] \;\; (N \times 1) \tag{13}$$

Hence using this method [22], the estimated global weighted vector calculated at the sink node is given as

$$w_k^{Glob} = w_{k-1}^{Glob} + \mu \sum_{i=1}^{M} \left( R_{DU,k} - R_{UU} w_{k-1} \right) \;\; (N \times 1) \tag{14}$$

According to instantaneous approximations [22], the global statistical information can be written as

$$\mathbb{R}_{UU,i} = u_i^T u_i \;\; (N \times N) \text{ and } \mathbb{R}_{DU,i} = u_i^T d_i \;\; (N \times 1) \text{ for } i \in M \tag{15}$$

Hence using instantaneous approximations, the global weighted vector $w_k^{Glob}$ (14) can be modified as

$$w_k^{Glob(IA)} = w_{k-1}^{Glob} + \mu\Big\{\Big(u_1^T d_1 - \big(u_1^T u_1\big) w_{k-1}^{Glob}\Big) + \Big(u_2^T d_2 - \big(u_2^T u_2\big) w_{k-1}^{Glob}\Big) + \\ + \ldots\ldots\ldots\ldots + \Big(u_M^T d_{M1} - \big(u_M^T u_M\big) w_{k-1}^{Glob}\Big)\Big\}$$

$$w_k^{Glob(IA)} = w_{k-1}^{Glob} + \mu \sum_{i=1}^{M} \Big(u_i^T d_i - \big(u_i^T u_i\big) w_{k-1}^{Glob}\Big) \quad (N \times 1)$$

$$\text{for} \quad i \in M \quad (16)$$

Similarly using LMS filter [22], the global weighted vector $w_k^{Glob}$ calculated at the sink node is given as

$$w_k^{Glob(LMS)} = w_{k-1}^{Glob} + \mu\Big\{u_1^T \Big(d_1 - u_1 w_{k-1}^{Glob}\Big) + u_2^T \Big(d_2 - u_2 w_{k-1}^{Glob}\Big) + .. \\ + \ldots\ldots\ldots\ldots + u_M^T \Big(d_{M1} - u_M w_{k-1}^{Glob}\Big)\Big\}$$

$$w_k^{Glob(LMS)} = w_{k-1}^{Glob} + \mu \sum_{i=1}^{M} u_i^T \Big(d_i - u_i w_{k-1}^{Glob}\Big) \quad (N \times 1)$$

$$\text{for} \quad i \in M \quad (17)$$

Now $w_k^{Glob(LMS)}$ is used to calculate the prediction filters (for $M$ nodes) at sink as $y_{Sink} = U w_k^{Glob(LMS)}$. This realizes that at this moment, the prediction filters at sink node are active to calculate $y_{Sink}$ and prediction filter $(y_i)$ used at each sensor node $i$ is still kept idle. Finally we use $y_{Sink}$ to calculate the prediction error at sink node as

$$error^{Glob} = [d - y_{Sink}] \quad (18)$$

We define a user defined error threshold $(\alpha)$ value to satisfy these two conditions defined as follows:

- If the $error^{Glob}$ (scalar value) calculated at the sink node for the respective node is greater than the error threshold $(\alpha)$ value, then the corresponding sensor node still continue to send the data to the sink node. This makes the filter at the sink node to adopt well for the received data transmitted form the corresponding node and goes to adaptive mode. In this situation, filter at the sink node (server) for this node is active and the filter for this corresponding node (client) is still kept ideal.
- But if the $error^{Glob}$ for the respective node is less than the error threshold $(\alpha)$ value, the global weighted vector $w_k^{Glob(LMS)}$ calculated at the sink node is transferred to the corresponding sensor node in the network. This transmission of $w_k^{Glob(LMS)}$ is like a *request* from the sink node to sensor node for stopping the transmission of data [23]. Once $w_k^{Glob(LMS)}$ is transferred from the sink node to sensor node, filter residing at sink node for the sensor node goes to ideal and it goes to prediction mode , finally filter residing for the corresponding sensor node is yet to become active.

***Phase III*** *Client:* Once $W_k^{Glob(LMS)}$ is received by each sensor node (conditioned *error* calculated at the sink node for the respective node is less than the threshold), it (client) uses $W_k^{Glob(LMS)}$ to calculate its new weighted vector, filter and error. Since $u_i$ observation is sensed by sensor node $i$, the desired signal scalar value calculated by each sensor node $i$ is given as

$$d_i^{new} = u_i w_k^{Glob(LMS)} + v_i \quad \text{where } i \in M \quad (19)$$

Hence the new updated weighted vector calculated at each sensor node $i$ in the network is given as

$$w_{i,k}^{new} = w_{i,k-1} + \mu\Big(u_i^T \Big(d_i^{new} - u_i w_{i,k-1}\Big)\Big) \quad (N \times 1)$$

$$\text{for} \quad i \in M \quad (20)$$

Now each sensor node finds its individual filter update value as $y_i^{new} = u_i w_{i,k}^{new}$ and finally the scalar error value calculated for each sensor node $i$ in the network is given as

$$error_i^{new} = \Big[d_i^{new} - y_i^{new}\Big] \quad (21)$$

Again using another user defined error threshold $(\beta)$ value, we illustrate two conditions:

- If $error_i^{new}$ calculate at the sensor node is greater than the error threshold value $(\beta)$, observation $u_i$ sensed by it is still transmitted to the sink node. At this time filter at node is active and goes to adaptive mode. At this stage, filter at the sink for the corresponding node is set idle.
- If $error_i^{new}$ calculate at the sensor node is less than the error threshold value $(\beta)$, the sensor node stop transmitting the observation $u_i$ to the sink node. Once data transmission is stopped, sensor node $i$ transfer it's $w_{i,k}^{new}$ to the sink node. This transmission of $w_{i,k}^{new}$ is like a *response* from the sensor node to the sink node that transmission of data is stopped. At this moment filter at the node is set ideal and goes to prediction mode. Sink node utilizes $w_{i,k}^{new}$ transmitted from each sensor node to track the signal statistics of each sensor node $i$ in the network. Once $error_i^{new} \geq \beta$, Repeat the same process as we explained in ***Phase I.***

Thus using these three phases of STDP model, we perform the *data reduction strategies* to reduce the communication overhead in the network. STDP model can track and learn the internal variation of the signal statistics without requiring a priori knowledge of the environment .Moreover it does the *joint prediction* scheme to capture spatial and temporal correlation of data among sensor nodes in the network.

### C. *Tracing Malicious Nodes in the Network*

Data gathering or data aggregation are the traditional procedure subjected to energy constraints [19] and maximizing [18] network life time. These procedures are done without verifying the data accuracy before data aggregation in the network. Hence without verifying the data accuracy before data aggregation cause problem if some of the senor nodes get malicious in the network. The sensor node gets malicious due to extreme physical environment e.g heavy rainfall or snow fall. Malicious nodes sense inaccurate data readings and transmit the inaccurate data to the sink node. Sink node aggregate inaccurate data with the other correct data send by the sensor nodes. Thus sink node estimate inaccurate data reading which results poor data gathering in the network.

In literature [7], we perform data accuracy for the network when some of the sensor nodes get malicious. Moreover we don't incorporate to trace and discard the malicious nodes from the network to get better data accuracy. Hence in this paper, we find a novel methodology to trace the malicious nodes in the network.

Since $M$ sensor nodes are randomly deployed over a sensor region, we assume some of the nodes get malicious due to extreme physical environment. But we don't know the extract number of malicious nodes out of $M$ sensor nodes in the network. Our motivation is to trace these malicious nodes in the network. The node is malicious or not depends upon the statistical behavior of observation $u_i$. We repeat the same procedure explained in STDP model of *Phases-I-III* where weighted vector $w_{i,k}^{new}$ gives the statistical information of each node to trace the malicious behavior. If the node is malicious, then the statistical value $(w_{mal_i,k}^{new})$ of that node is different from the normal [7] nodes .Finally we can trace the malicious nodes and discard it from the network to get better data accuracy and data aggregation under spatially correlated data.

## IV. Evaluation

To perform the simulations, a sensing region of $4m \times 4m$ grid based wireless sensor topology is taken with a sink node in the network. We deploy ten sensor nodes in the sensing region. Each sensor node can sense the observations (e.g temperature) from the physical environment and transmit it to the sink node. Since our system model has the ability to trace the internal variations of the signal statistics to adopt itself with the physical environment, we consider such a tropical environment for our experiment where the variations of signal statistics (e.g temperature) are much more for certain duration. Our system model can work well to trace the variations of signal statistics for tropical desert area like Jaisalmer (Rajasthan-India). On 26th January 2012, the minimum and maximum temperatures recorded in jaisalmer are $7^{\circ}$Celsius and $22^{\circ}$Celsius respectively according to [28]. Such variations of temperature for a particular duration are the subject of interest to measure the variation of signal statistics of temperature rather than measuring the temperature variation of room temperature using our system model. Since the temperature variation of room temperature is very less say for example $26^{\circ}$Celsius to $30^{\circ}$ Celsius for a particular duration. Hence for our all simulation purposes, we generate random data (temperature) using matlab which is sensed by sensor nodes to validate our results for ADA model and STDP model respectively. Thus the sensor nodes reported random (temperature) data once every 15 minutes recorded over one day (26th Jan 2012) on jaisalmer . Another example is to choose subtropical highland climate like Mawsynram[1] and cherrapungi ,(Meghalaya- India) where our system model can works well to trace the variations of the signal statistics (measure the rainfall in mm) in the sensing region.

In the first simulation setup, we estimate the data accuracy of the signal statistics at the sink node for all the deployed ten sensor nodes in the sensing region using ADA model. Fig 3(a) shows how the accuracy of the data collected at the sink node converges for the ADA model and the data accuracy model [8, 9] with respect to time (iterations). In Fig 3(b), we perform data accuracy of the signal statistics at the sink node for the ADA model with respect to number of sensor nodes in the network. Simulation results shows that about six sensor nodes are sufficient to perform approximately the same data accuracy as achieve by the ten sensor nodes in the sensing region. Thus an optimal (six) sensor nodes can perform data accuracy using ADA model instead of using ten sensor nodes in the network. We choose about six sensor nodes which are almost close to the sink node are eligible to perform the data transmission in the network maintaining a desired accuracy level using ADA model. Thus reducing the number of sensor nodes for data transmission maintaining a desired accuracy in the network can reduce the communication overhead and increase the lifetime of the network.

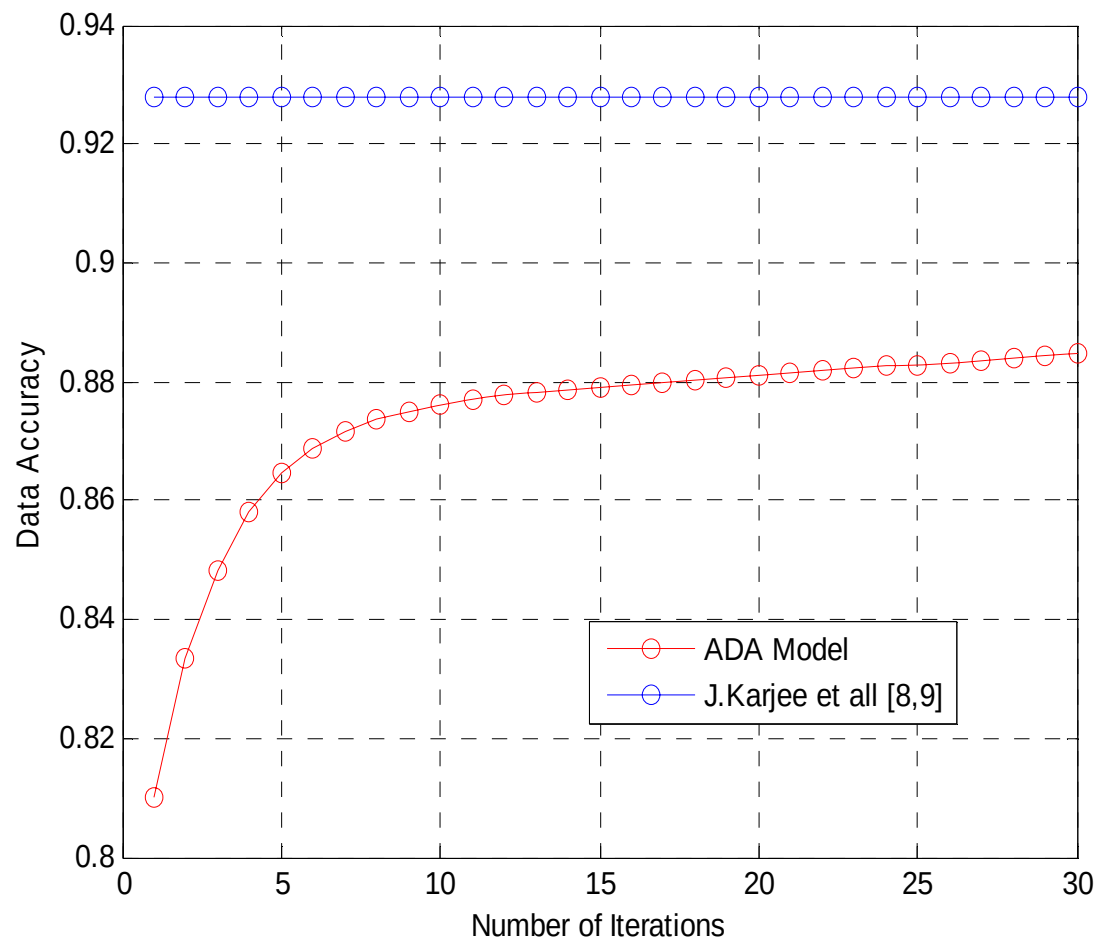

Fig 3 (a) Plot for data accuracy versus number of iterations.

[1] According to [29],Mawsynram is considered as the wettest place in the world with highest average annual rainfall

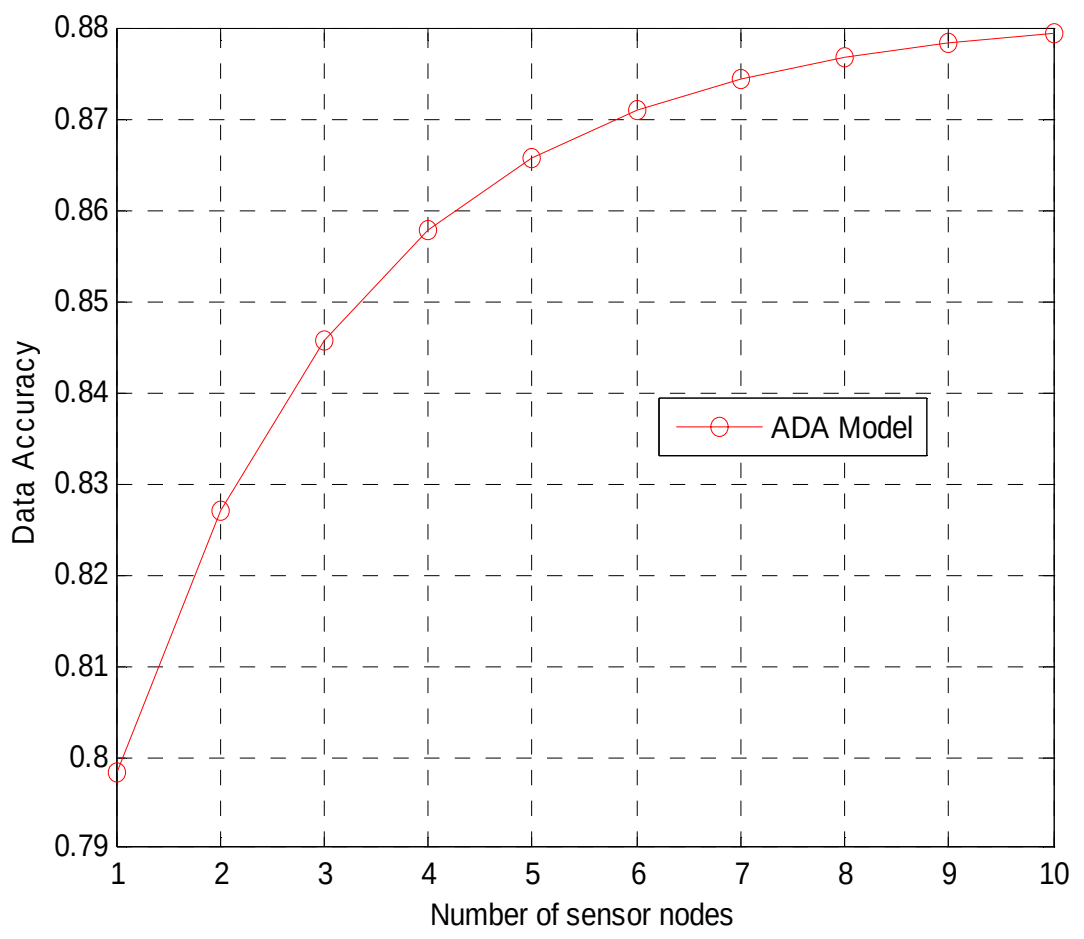

Fig 3(b) Plot for data accuracy versus number of sensor nodes

In the second simulation setup, we illustrate the performance analysis of STDP model. Since using ADA model, we select about six sensor nodes (optimal sensor nodes selected ) to perform the data transmission instead of using ten sensor nodes, we assume these sensor nodes are with node Id's 2,4,5,7,9 and 10 selected for data transmission of signal statistics. These sensor nodes are chosen such that they are close to the sink node in the sensing region. Further using STDP, we can reduce the communication overhead of the data transmission for these six sensor nodes to the sink node in the network. To analysis our simulation results in Fig 4(a) using STDP model, we report data transmission percentage of sensor readings for optimal number of nodes selected with respect to a varying error threshold value $(\beta)$. For each sensor node, a prediction error is calculated for each data stream of ( $N$ block) sensor readings to transmit. If the prediction error is greater than the error threshold value, the respective data stream of ( $N$ block) sensor readings are transmitted by each sensor node to the sink node. Instead of transmitting all the data streams, a subset of data stream sensor readings for each sensor node is delivered to the sink node using data reduction strategies. Moreover Fig 4(a) also shows statistical variations of signal for sensor nodes in the network. The statistical data streams among sensor nodes are almost similar because the data streams are spatially and temporally correlated among them. In Fig 4(b), we compare the percentage of data transmission for data stream block size $N = 4$ and data stream block size $N = 5$ for sensor node Id-2 with respect to error threshold value $(\beta)$. The simulation result shows that if we transmit data stream block size of $N = 4$, we can reduce the percentage of transmission cost effectively than transmitting data stream block size of $N = 5$. Another conclusion is drawn as the data stream block size $(N)$ is small, tracking and learning of statistical signal is easier whereas if we use data stream block size larger, a better estimation of statistical is performed.

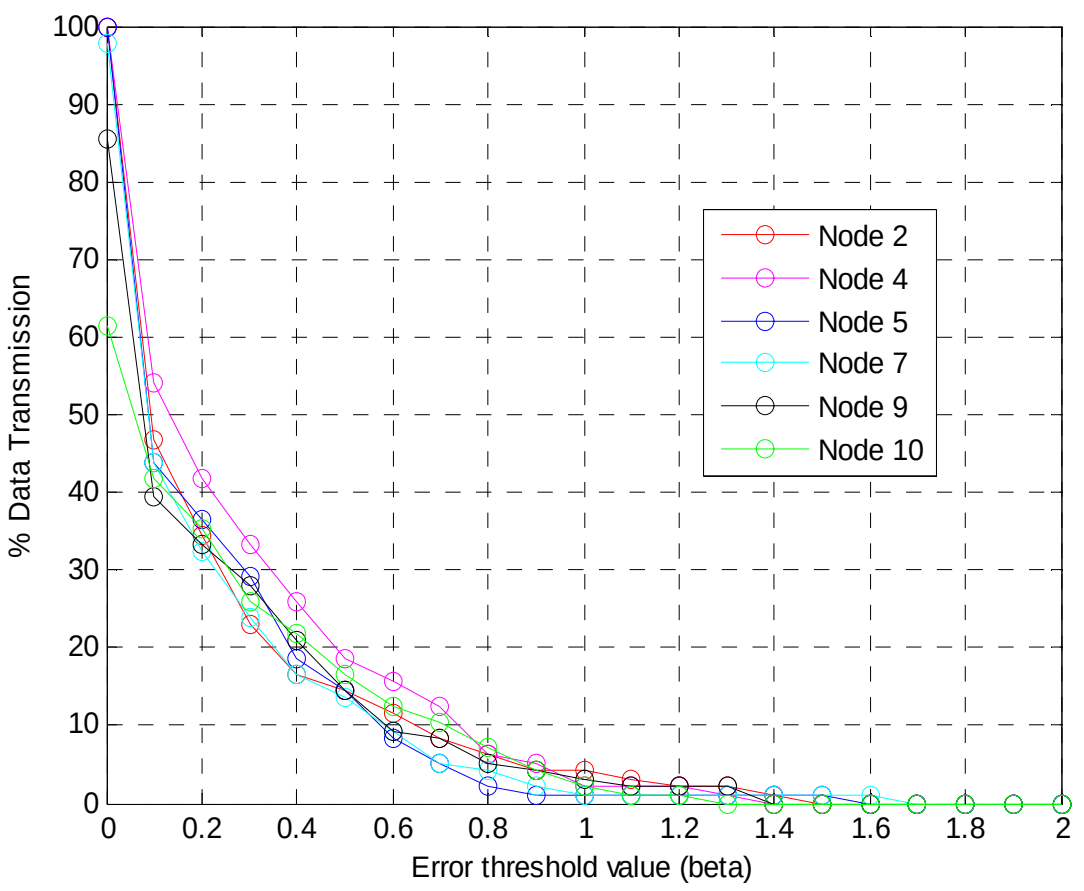

Fig4 (a) Percentage of data transmission versus error threshold value $(\beta)$ .

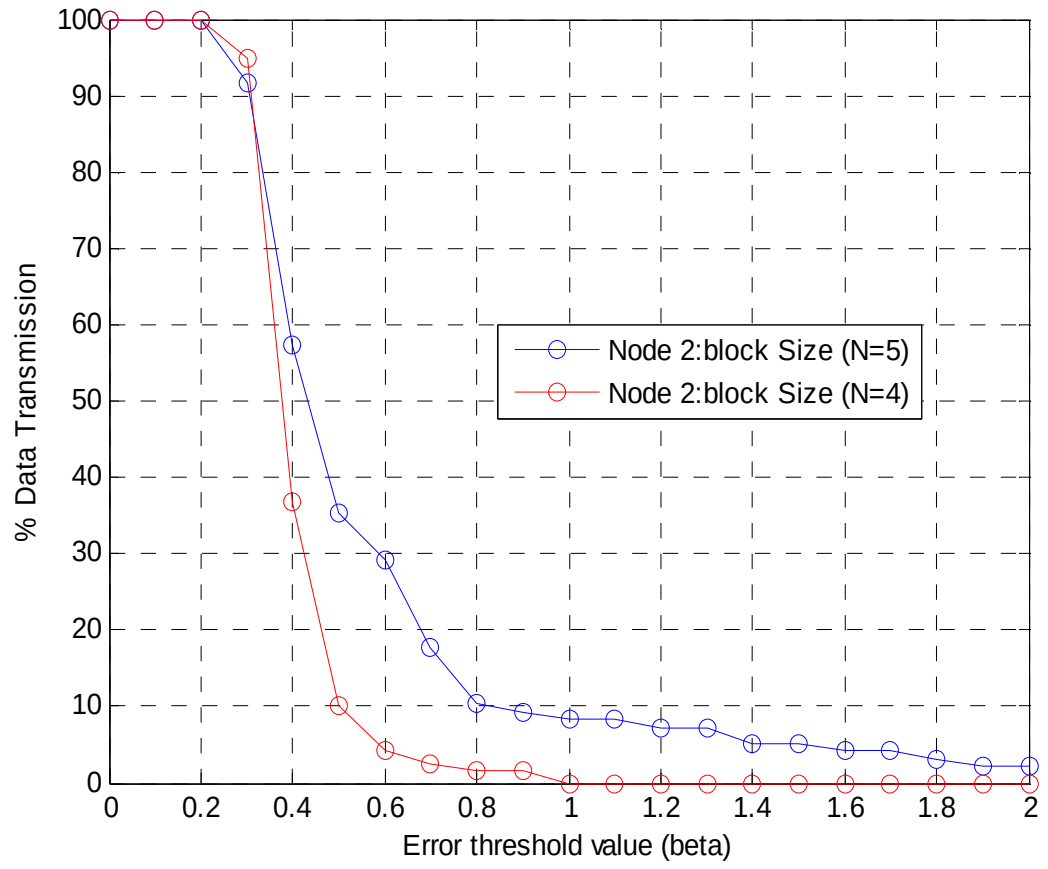

Fig4 (b) Percentage of data transmission versus error threshold value $(\beta)$ for Block size $N = 4$ and $N = 5$ data streams respectively.

Finally in the third simulation setup using STDP model, we can find the number of malicious nodes in the network. Malicious node can sense inaccurate data readings .The signal variations of malicious nodes are much higher than normal nodes. In our network, assuming node Id's 5 and 9 are malicious and node Id's 2, 4, 7 and 10 are normal nodes. Fig 5 shows that variations of the weighted vector of node Id's 5 and 9 which are much higher than the normal nodes. The weighted vector $(w^{new}_{mal_i,k})$ of malicious nodes shows abnormal signal variations than the normal nodes. Such abnormal signal variation of the weighted vector of sensor node detected at the sink node is said to be malicious node. Thus we can easily trace the number of malicious nodes in the network by analyzing the signal variation of the weighted vector ( $w^{new}_{i,k}$ ) of each node. The signal variation of weighted vector and the variance of each sensor node are summarizes in Table I. Thus from Table I , we conclude that the weighted vector of node Id's 5 and 9 shows abnormal statistical variations than the normal nodes. Finally node Id's 5 and 9 can be discarded from the network to get better data accuracy in the network.

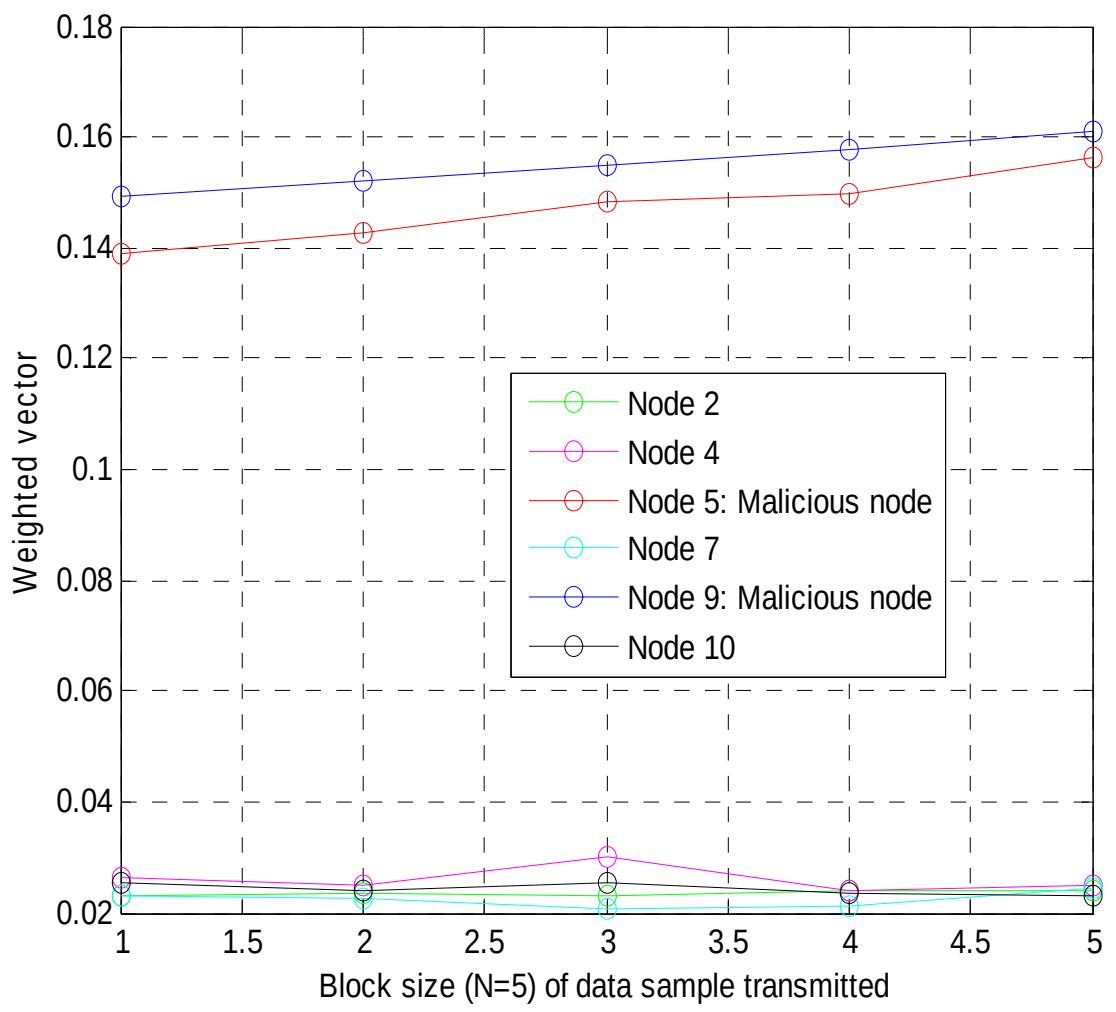

Fig 5. Weighted vector versus data sample block size (N=5) of each sensor node in STDP model to trace malicious nodes.

| Selecting Nodes Using ADA | Node 2 | Node 4 | Node 5 | Node 7 | Node 9 | Node 10 |
|---|---|---|---|---|---|---|
| Weighted vector of sensor nodes to show the statistical behaviour using STDP model | 0.0232<br>0.0236<br>0.0230<br>0.0240<br>0.0238 | 0.0264<br>0.0248<br>0.0302<br>0.0242<br>0.0250 | 0.1390<br>0.1429<br>0.1484<br>0.1500<br>0.1561 | 0.0229<br>0.0225<br>0.0207<br>0.0214<br>0.024 | 0.1493<br>0.1520<br>0.1551<br>0.1575<br>0.1609 | 0.0253<br>0.0242<br>0.0253<br>0.0236<br>0.0230 |
| variance | 1.3760 e-07 | 4.6816 e-06 | 3.4838 e-05 | 1.6296 e-06 | 1.6538 e-05 | 8.3760 e-07 |
| Result | *Nor* | *Nor* | *Mal* | *Nor* | *Mal* | *Nor* |

Table I: Weighted vector of sensor nodes to trace malicious nodes in the network.( *Nor:* Normal nodes and *Mal:* Malicious nodes)

## V. CONCLUSIONS

In this paper, we presented Adaptive Data Accuracy (ADA) model to select an optimal number of sensor nodes in the network under adaptive approach. ADA model doesn't require any a priori knowledge of the signal statistics of the environment. Moreover we describe Spatio-Temporal Data Prediction (STDP) model which reduces the communication overhead for these optimal sensor nodes under data reduction strategies. Simulation results show that STDP can learn and track the internal variation of signal statistics of the environment. Finally we propose a mechanism using STDP model to trace the malicious nodes in the network if any, due to extreme physical environment e.g heavy rainfall. Extensive simulation results are performed to validate ADA and STDP models respectively under malicious network.